\documentstyle[12pt]{article}
\newcommand{\be}{\begin{equation}}
\newcommand{\ee}{\end{equation}}
\newcommand{\ba}{\begin{eqnarray}}
\newcommand{\ea}{\end{eqnarray}}
\newcommand{\lb}{\label}

\begin{document}
\begin{center}
\huge Chaos in Black Holes Surrounded  by Gravitational Waves
 
\end{center}

\vspace{3ex} 

\centerline{  \em P.S. Letelier\footnote{e-mail: letelier@ime.unicamp.br}
 and 
W. M. Vieira\footnote{e-mail: vieira@ime.unicamp.br} } 
\vspace{1ex}

\begin{center}
 Departamento de Matem\'atica Aplicada-IMECC\\
Universidade Estadual de Campinas\\
13081-970 Campinas,  S.P., Brazil \\ \vspace*{0.3cm}{ \small 

September, 1996}

\end{center}

\baselineskip 0.7cm
 \vspace*{6ex}
 The occurrence of chaos for  test particles 
moving  around  Schwarzschild black holes perturbed by a special 
class of gravitational waves is studied in the context of the Melnikov 
method. The  explicit integration of the equations of motion  for the
 homoclinic orbit is used to reduce
 the application  of this method to the study of simple graphics.\\

\newpage
\section{Introduction}

\hspace{3em}
There are two main lines of research of chaotic behavior in General
 Relativity: one deals with chaoticity associated to inhomogeneous
 cosmological models, as in Bianchi IX model \cite{BKL}.  
The invariance  of the relativistic theory  challenges the proper
 concept of chaos in 
this approach \cite{gemat}.  Although, a gauge invariant
 method based  on Maupertuis principle is proposed in~\cite{Szy2}
and applied with little changes
to the Bianchi IX  model~\cite{Biesiada}.
 The method points to the occurrence
of local instability in
this cosmological model near its singularity. This adds to a
plethora of early analysis and numerics
 (see for instance the references cited in~\cite{Biesiada}), which 
 reinforces  the increasingly accepted idea that
 Bianchi IX is chaotic in some meaningful sense.

 The other line assumes a given geometry and looks for chaotic behavior
 of geodesic motion in this background. In this case, the geometry 
can be taken as either an approximate 
or an exact solution of Einstein equations. Examples of chaotic
 geodesic motion are considered in \cite{cont} -- \cite{vielet1}
 (exact geometry) and  in \cite{bc,moe} (approximate). 

 Due to its universality and 
 intrinsic mathematical interest, models in which unstable
periodic  orbits (UPOs) are subjected to  small periodic perturbations  
has been one
 of the main paradigms of deterministic chaos \cite{arnIII}. An 
  analytical tool
to study these models is the Melnikov function that describes the 
transversal distance between the unstable and stable manifolds 
emanating from an UPO. Its isolated odd zeros indicate the crossing
 of these manifolds, hence the onset of chaos \cite{mel,hol}.  
Examples of applications of the Melnikov method in gravitation
 are: the gravitational collapse of cosmological 
 models \cite{ivano}, the
 study of orbits around a black hole perturbed by either gravitational
 radiation \cite{bc} or an external quadrupolar shell \cite{moe}. Also
 for particles moving in several models of attractive 
centers periodically perturbed this method has been
 applied \cite{letvie1}. 

In this work we study the occurrence of chaos for  test particles 
moving  around a Schwarzschild black hole perturbed by a special 
class of gravitational waves  \cite{xan}. Since
 this perturbation is an exact solution of the Regge-Wheeler
 equations (see for instance \cite{chandra})  we can study the range 
of the parameters of the model in order to have chaos. In \cite{bc} a
 similar situation is studied but the perturbations, albeit more
 general that ours, are known only in the high frequency limit.

In the next section we present a summary of the Melnikov method. In 
Sec. 3 we review the homoclinic orbits for Schwarzschild solution, and
present an explicit  solution $t=t(r)$ for this orbit. In sections
4 and 5 we consider the perturbations and the 
the  application of the Melnikov method, respectively.
 In the last section we 
discuss  some of the previous results and also compare our work
 with the application of the Melnikov method presented in \cite{bc}.

 \section{The Melnikov method}

\hspace{3em}
Let us consider a Hamiltonian,
\begin{equation}
H_0=\frac{p^2}{2m}+ V(q) , \label{H0}
\end{equation}
that admits at least one UPO with the corresponding homoclinic  
 orbit,  and also a small periodic
 perturbation described  by the Hamiltonian function
 $\epsilon H_1(q,p,t)$. Then the transversal distance, in phase
 space, between the perturbed 
 unstable  and the perturbed stable manifolds emanating
 from the UPO is proportional to \cite{hol}
\begin{equation}
M(t_0)=\int_{ -\infty}^{ \infty} {\{ H_0,H_1\}dt}, \label{M}
\end{equation}
where the integral is 
 taken along the {\em unperturbed}  homoclinic orbit, $t_0$ is
 an arbitrary initial time, and
\begin{equation}
\{f,g\}=\frac{\partial f}{\partial p}\frac{\partial g}{\partial q}
-\frac{\partial g}{\partial q}\frac{\partial f}{\partial p}, \label{pb}
\end{equation}
with $(q,p)$ being canonical conjugated variables. If  there is an
isolated  intersection for some $t_0$, i.e., an isolated odd  zero of
 $M(t_0)$, then there will be one for every  $t_0$. This infinitely
 many crossings of manifolds   will produce a tangle that is the 
signature of the homoclinic chaos \cite{mel,hol}. 
    
In particular, if $H_1$ has the form
\be
H_1(p,q,t)=\hat{H}_1(p,q) \cos(\omega t)  , \lb{H1}
\ee
we have \cite{hol}
\be
M(t_0)= \cos(\omega t_0 )\int_{ -\infty}^{ \infty} {\{ H_0,\hat
 H_1\}\cos(\omega t)dt}+ \sin(\omega t_0 )\int_{ -\infty}^{ \infty} 
{\{ H_0,\hat H_1\}\sin(\omega t)dt} \lb{M2}
\ee
Then, in the generic case, we will have isolated zeros when  at
 least one of the integrals of the previous formula is different
 from zero.

\section{The homoclinic orbits}

We shall consider a relativistic particle 
moving in  a fix
spacetime described by the metric $g_{ab}$. The world-line of a 
particle will be denoted by $x^a(s)$ and its mass by $\mu$. The motion
 of the particle can be obtained  from the action
\be
S[x]=\frac{\mu}{2}\int g_{ab} \dot{x}^a \dot{x}^b d s \lb{S}.
\ee
This action is not reparametrization invariant, but is the simpler
 for our purposes. $s$ is the  proper time along the world-line. The 
canonical conjugate momentum to $x^a$ is $p_a=\mu g_{ab} \dot{x}^b$ 
and satisfies the mass shell constraint $g^{ab}p_a p_b = -\mu^2$. The
 Hamiltonian of the system is
\be
H=\frac{1}{2\mu} g^{ab} p_a p_b. \lb{Hg}
\ee
The background metric will be considered as the metric of a
 non rotating black hole, i.e., the Schwarzschild metric,
\be
ds^2=-fdT^2+f^{-1}dR^2+R^2(d\vartheta^2+\sin^2\vartheta 
d\varphi^2) ,  \lb{ds}
\ee
where $f=1-2M/R$. Since, $T$ and $\varphi$ are cyclic variables we have
the conserved quantities:
\ba
&&E\equiv -p_T =\mu f \frac{dt}{ds} \nonumber \\
&&L\equiv p_\varphi=\mu R^2 \sin^2 
\vartheta \frac{d \varphi}{ds}.\lb{EL}
\ea
For planar motion $\vartheta=\pi/2$ ($\dot{p}_\vartheta
 =p_\vartheta =0$) we have the
 equivalent one dimensional problem of a particle of mass $2\mu^2$ moving
in the the potential $V$ and constant energy $E^2-\mu^2$,
\ba
&& \mu^2 \left(\frac{dR}{ds} \right)^2 +V(R)=E^2-\mu^2 \nonumber\\
&& V(R)=-\frac{2M\mu^2}{R}+\frac{L^2}{R^2}-\frac{2ML^2}{R^3} . \lb{V}
\ea
It is convenient to work in dimensionless variables. Defining $r=R/2M$
and $U=(2M/L)^2 V$ we find
\be
U = -\frac{(1-\beta^2)}{3r} +\frac{1}{r^2} -\frac{1}{r^3} ,\lb{U}
\ee
where $\beta=\sqrt{1-12M^2\mu^2/L^2}<1$. Since  $U(\infty)
 \sim -{\sl O}(1/r)$,  in order to have a local maximum (or unstable point, 
$r=r_u$) with $U(r_u)<0$,   we need $0\leq \beta \leq 1/2$. 

In Fig.1 we plot $U$ as a function of $r=R/2M$ for various values
 of $\beta$. The  unstable periodic orbit (UPO) correspond in phase
 space to the point $(r,p_r)=(r_u,0)$. Now by taking 
 $U(r_u)=(2M/L)^2(E^2-\mu^2)$ we
 get the motion equation for the homoclinic orbit,
\be
\frac{dr}{dt}=\pm\frac{ w_\beta(r-1)(r-r_u)(r_m-r)}{r^{3/2}}, \lb{me}
\ee
where  $r_m=3/(1-2\beta)$, and   $w_\beta=\frac{3}{2}\sqrt{3(1
-2\beta}/(2-\beta)^2$. We have used the constant of motion $E$ 
to change the parameter of evolution $s$ to the dimensionless
 time coordinate
$t=T/2M$. So the homoclinic orbit is limited
 by $r=r_u$ and the maximum value $r=r_m$ (turning point).
 This  motion equation
admits the primitive, 
\ba
&&\pm w_\beta t =[r(r_m-r)]^{1/2}/r_m+(2+2r_u+
r_m)\arctan\sqrt{(r_m-r)/r}\nonumber\\
& & \hspace*{2cm}-\frac{2r_u^{5/3}}{(1-r_u)\sqrt{r_m-r_u} 
}{\rm arctanh}\sqrt{\frac{r_m/r-1}{r_m/r_u-1}} \nonumber\\
&&\hspace*{2cm}+ \frac{2}{(1-r_u)\sqrt{r_m-1}}{\rm 
arctanh}\sqrt{\frac{r_m-r}{r(r_m-1)} } . \lb{t}
\ea
We have chosen the constant of integration to have $t(r=r_m)=0$.
In Fig.2 we show the positive branch of  (\ref{t}) for different values of
$\beta$. The particle takes an infinite amount of time to depart 
(arrive ) from (to) the homoclinic point where the UPO is 
located, this is a universal behaviour for these type of orbits. 
The explicit expression (\ref{t}) for the homoclinic orbit
 will play an important role in 
our study of the Melnikov function and its  use 
 will be a  significant departure from  the treatment
given in \cite{bc} wherein other interesting graphics related with
the homoclinic orbit are presented.  

\section{The perturbations}

The perturbations of the black hole that we shall consider are 
of the particular class $g_{ab}+\epsilon h_{ab}$ with $g_{ab}$ given
by (\ref{ds}) and 
\ba
&&h_{TT}=- fX P_l \cos(\sigma T) \nonumber \\
&& h_{RR}=f^{-1}YP_l\cos(\sigma T)\nonumber\\
&& h_{\vartheta \vartheta}=R^2(ZP_l+W P_l,_\vartheta, _\vartheta 
)\cos(\sigma T)\nonumber\\
&& h_{\varphi \varphi}=R^2\sin^2\vartheta(Z P_l +
WP_l,_\vartheta \cot\vartheta)\cos(\sigma T) ,\lb{h}
\ea
where $X,Y,Z$ and $W$ are functions of $R$ to be determined by the
 Einstein equations $R_{ab}(g_{cd}+\epsilon h_{cd})=0.$
 $P_l=P_l(\cos\vartheta)$ are the usual Legendre polynomials. When 
the above
mentioned Einstein equations are expanded up to
the  first order  in $\epsilon$ we find  a linear system of
first order differential equations for the functions $X,Y,Z$ and $W$
\cite{chandra}. The particular separation of the angular part of
  the perturbation (\ref{h}) was found by Friedman \cite{frie}.
 Combinations of this variables  reduce these equations
to the Zerilli equation, i.e., to a Schr\"odinger type of equation with
a nontrivial potential \cite{chandra}.  A particular solution 
of the first order differential equations is 
\be
X=pq, \;\; Y=3 M q, \;\; Z=(R-3M) q, \;\; W=R q, \lb{x1}
\ee
where
\be
q=f^{1/2}/R^2, \;\; p=M - \frac{M^2+\sigma^2R^4}{R-2M}. \lb{x2}
\ee
This solution was found by Xanthopoulos \cite{xan} using an {\em ad
 hoc } method. We note that it can be easily found in the 
 Zerilli formulation and  corresponds to zero Zerilli function $Z^{(+)}$,
\be
Z^{(+) }=\frac{R^2}{nR+3M}(\frac{3M}{R}W -Y), \lb{z} 
 \ee
where $n=(l-1)(l+2)/2$.

From  Eq.(\ref{Hg}) we have for the perturbed system
\be
(g^{ab}-\epsilon h^{ab})p_a p_b= -\mu^2. \lb{pp}
\ee
For an even $l$ we have that the perturbations (\ref{h}) are even functions
in the variable $\cos{\vartheta}$, i.e., we have perturbations with
reflection symmetry with respect the plane $\vartheta=\pi/2$.
Then, for a particle moving in the plane $\vartheta=\pi/2$ these
type of perturbations do not change the plane of  the orbit. In other  words, 
 the perturbation introduces a  ``perpendicular force"
 that is an odd function of $\cos{\vartheta}$. 
We find for particles moving in the plane $\vartheta=\pi/2$,
\be
-p_t=H_0 +\epsilon H_1 , \lb{pp2}
\ee
with
\be
H_1=-[ \frac{E}{2f}h_{TT} -
\frac{1}{2E}(f^3p_R^2 h_{RR}+\frac{L^2f}{R^4}h_{\varphi\varphi}) 
] ,\lb{H12} 
\ee
and 
\ba 
&&h_{TT}= -fpqN_l\cos(\sigma T), \nonumber\\
&& h_{RR}=\frac{3Mq}{f}N_l\cos(\sigma T), \nonumber\\
&&h_{\varphi \varphi}=R^2(R-3M)qN_l\cos(\sigma T), \lb{hh}
\ea
where  $N_l
$ is zero for odd $l$ and  
\be
N_l=(-1)^n \frac{1 \cdot 3 \cdot\cdot\cdot (2n-1)}{2\cdot
 4\cdot\cdot\cdot 2n}, \;\; (l=2n) .\lb{N}
\ee
The order of Legendre polynomial $l$ enters in a trivial way in this 
class of planar perturbations, as a multiplication 
by the constant $N_l$. The
frequency $\sigma$ not only appears in the oscillating  function   
 $ \cos(\sigma T)$ but also in  $p$.

\section{The Melnikov function}

Now we shall use the fact that we know explicitly the function $t=t(r)$
for the homoclinic orbit to evaluate the Melnikov function. We find
\be
M(t_0)=-\sin(\omega t_0) K,\lb{MM}
\ee
with  
\ba
&&K=\frac{E}{M} N_l J ,\lb{K}\\
 && J=\int^{r_m}_{ru}F(r,\beta)  \sin(\omega t(r))dr, \lb{j} 
\ea
where $\omega =2M\sigma$ is a dimensionless frequency and
\ba
F(r,\beta,\omega )=\left[ \frac{1}{2r^2}\hat{h}_{tt}-
\frac{f}{2}h_{tt,r} +
\frac{2\gamma f^2}{r^5}(f-\frac{1}{4r})\frac{ \hat{h}_{\varphi\varphi} 
}{(2M)^2}  -\frac{\gamma f^5}{2r^4} \frac{ \hat{h}_{\varphi\varphi ,
 r} }{(2M)^2}  \right.  
 \nonumber \\
 +\frac{f^2}{2r^2}(1-2\gamma f^2)\hat{h}_{rr}
 -\left( \frac{dr}{dt}\right)^2\left(\frac{1}{r^2}\hat{h}_{rr} +
\left. \frac{f}{2} \hat{h}_{rr,r}\right)  \right]. 
\ea
The constant $\gamma $ is defined in terms of $\beta$ as  $\gamma=
\frac{2}{27}(1+\beta)(2-\beta)^2$, and
\ba
&&\hat{h}_{tt}=-\frac{f^{3/2}}{2r^2}\left( 1 -\frac{1+4\omega
 r^4}{2rf}\right),\nonumber\\
&&\hat{h}_{rr}=\frac{3}{2r^2f^{1/2}},\nonumber\\
&&\frac{\hat{h}_{\varphi\varphi}}{(2M)^2} = f^{1/2}(2r-3)/2. \lb{hhat}
\ea
By $t=t(r)$ in (\ref{j}) we mean the positive branch of (\ref{t}). 
We have used the homoclinic orbit to map the infinite interval of the 
integration  to a finite one. Also, all the quantities appearing in 
the definition of $J$ are dimensionless.

Therefore,  to have the homoclinic tangle of orbits we need $J\not=0$.
The  exact computation of $J$ even  in the simple case of the  perturbations 
under consideration is hopeless. Nevertheless, the integrand of $J$ is 
explicitly known
in terms of the variable of integration. This fact can be used 
 to study the zeros of $J=J(\beta,\omega)$, or better the range of
 $\beta$ and $\omega$ in which $J\not=0$,  in a simple graphic way.

Firstly,  it is instructive to plot the function $S=\sin(\omega t(r))$ for
a fix $\beta$ and different values of $\omega$.  We find (see Fig.3)
   that,  depending on the  value of $\omega$,  $ S$ does not
 change  sign in a large
 portion of the interval  $r_u(\beta)\leq r \leq
 r_m(\beta)$, e.g.,  the two bottom 
curves (from right to left). We also have a very rapid oscillation near 
 the position of the UPO, that is not shown in 
the graphic. The values of 
the parameters  are $\omega=0.15$ (top, from right
 to left), 0.1, and 0.05 (bottom)  and $\beta=1/4$. Also it
is illuminating to make a graphic of $F(r,\beta,\omega)$ for different
 values of $\beta$ and $\omega$ in
 the range of interest $r_u(\beta)\leq r
 \leq r_m(\beta)$. Fig. 4 shows that this function is very 
well behaved and
does not change of sign in the interval of interest. The integrand
of $J$, i.e.,  $ F(r,\beta,\omega)\sin(\omega t(r))$ with $\beta =1/4$ and
$\omega$=0.1 is presented in Fig.5. The area under the
 curve is clearly
different from zero, then $J\not =0$. In this case the
 perturbation will 
originate a chaotic motion of the test particle.  The value 
$\omega=0.1$ was chosen from the relation
\be
 \omega_k\sim  \pi/t_k , \lb{k}
\ee
where $t_k$ is the value of the  homoclinic orbit that correspond
 to  the maximum value of its curvature $k=t''/(1+(t')^2)^{3/2}$.
For $\beta=1/4$ we have $t_k$=29.3691 and $\omega_k$=0.1069.
Therefore, we can say that the perturbations with  $\omega < \omega_k$  
will originate chaotic behaviour.

\section{Discussion}

To better understand the particular perturbation (\ref{h}) we have 
computed the Kretschmann scalar for the  perturbed metric for the
 particular values of $l=2$ and $l=3$, we find
\be
R^{abcd}R_{abcd}=\frac{48M^2}{R^6} + \epsilon k_l \frac{M}{R^8}(3M
-R)\sqrt{1-\frac{2M}{R}}P_l (\cos\vartheta ),\lb{kr}
\ee
the numerical  constant $k_l$ is $k_2=58$ and $k_3=-3/2$. Thus, we 
have a  Kretschmann scalar  decaying as $R^{-7}$ for
$R \gg 2M$, indicating that part of the wave has origin
 in the black hole.

In this work we have shown that for a particular
 class of gravitational  perturbations we have that
the particles moving along timelike geodesics 
will present a chaotic motion when
the frequency of the perturbation  is $\omega<\omega_k$. In the
 complementary case,  $\omega>\omega_k$ our simple graphic method does
 not apply. But, anyway  one can compute numerically 
the integral $J$  with a great precision ( the integrand 
 is known explicitly).  
We found  that for $\omega>\omega_k$,  $J$  has  positive 
and negative values  depending on the  values of $\beta$.  Then  for 
certain specific values of $\beta$ we will have  $J(\beta,\omega)=0$ 
and the Melnikov method does not apply to these cases.
From the previous analysis it does look like that the generic
situation is chaotic, i.e., only for a numerable set of
frequencies $\omega$,   given a value of $\beta$,
  we should  not have chaos. The number of elements of this
 set increases with  $\omega$, but is still numerable.
To be more precise, the Melnikov method fails to 
predict chaos only for a a numerable set of values 
of the parameters.
  We note that seldom one can  find so easily  the
 parameter  range  of a given problem
that will produce a chaotic situation. Moreover,
for parameters outside this  range  it is reasonable to guess that
only  for a numerable set of frequencies the integrability is
 preserved. Therefore, based in our
 previous analysis,  we can conjecture that chaos is generic for the 
class of perturbed systems studied.
In our case, the limit $\lim_{\omega \rightarrow\infty}J(\beta,\omega)$
is not defined. Note that in $F(\beta,\omega)$  appears  $\omega^4$
and in the integrand we also have $\cos{\omega t}$.
 In the case analyzed in \cite{bc} the perturbation are 
quite general, but
 they are known explicitly only in the limit
 $ \omega \rightarrow\infty$. Their  integral equivalent to our $J$  has
 $\pm\infty$ integration 
limits and in the integrand they have, essentially,  an 
oscillating  function
 with frequency $\omega$ multiplied by another function proportional
 to $\omega$.  Since, in \cite{bc},  every thing is known only in the 
limit of infinite
 frequency, in order to predict chaotic behaviour, it is necessary to
show that their integral  is different from zero in the same limit. 
This point is unclear in the quoted paper.

The authors thank CNPq and FAPESP for financial support and 
Prof. D.V. Gal'tsov (Moscow) for enlightening discussions about 
black hole perturbations.

\newpage
\noindent
FIGURE CAPTIONS

\vspace*{1cm}
\noindent
Fig.1.    The effective $U$ potential is plotted for the dimensionless
 parameter   $\beta=$ 1/3 (top curve), 1/4,
 and 1/5 (bottom curve); the maximum value is located at 
 $r_{un}=3/(1+\beta)$.\\

\vspace*{0.4cm}
\noindent
Fig.2.  A graphic of the positive branch of Eq. (\ref{t}) 
is pictured  for  values of the
 parameter $\beta= $ 1/3 (top curve), 1/4,  and 1/5 (bottom curve). We 
have that the particle takes a finite time to travel
 from $r_m$ to the vicinity of
 $r_{un}$ and then takes an infinite time to arrive to 
$r_{un}$ itself, wherein is located the UPO.\\
 
\vspace*{0.4cm}
\noindent
Fig.3.  A graphic of \, $S=\sin [\omega\tau(r)] $
 for $\beta=1/4$, and $\omega =$0.05 (bottom  curve, from right 
to left), 0.1, and 0.15 (top curve). \\

\vspace*{0.4cm}
\noindent Fig.4.   A plot of $F(r,\beta,\omega) $  for $\beta$=1/4.

\vspace*{0.4cm}
\noindent Fig.5.  A  plot of $F(r,\beta,\omega)\sin(\omega t(r)) $ for 
$\beta$=1/4, and $\omega$= 0.1. We have a rapid
 oscillation near $r_{un}$ (=12/5 )  (not  shown in the figure),
that produce very fine spines in the range  [-0.03, 0.03].


\newpage
\begin{thebibliography}{99}

\bibitem{BKL} V.A. Belinskii, I.M. Khalatnikov and E.M. Lifshitz, 
Adv. Phys.  19, 535 (1970).
\bibitem{gemat} G. Francisco and G. E. A. Matsas, Gen. Relat.
 Grav.  20, 1047 (1988).
\bibitem{Szy2}
M. Szyd\l owski, Phys. Lett. A 176, 22 (1993).
\bibitem{Biesiada}
M.  Biesiada, Class. Quantum Grav. 12,
 715 (1995).
\bibitem{cont} G. Contopoulos, Proc. R. Soc. Lond. A 431, 183 
(1990); A 435, 551 (1991).
\bibitem{ssmae} Y. Sota, S. Suzuki and K. Maeda, Class. Quantum 
Grav. 13, 1241 (1996). See also, W.M. Vieira and P.S. Letelier,
Class. Quantum Grav. 13, 3115 (1996).
\bibitem{karvo}
V. Karas and D. Vokrouhlick\'y, Gen. Rel. Grav. 24, 729 (1992).
\bibitem{vielet1} W.M. Vieira and P.S. Letelier, Phys. Rev. 
Lett. 76, 1409 (1996); Errata, 76, 4098 (1996).
\bibitem{bc} L. Bombelli and E. Calzetta, Class. Quantum Grav. 
9, 2573 (1992).
\bibitem{moe} R. Moeckel, Comm. Math. Phys. 15, 415 (1992).
 \bibitem{arnIII} V.I. Arnold, ``Dynamical systems (Encyclopaedia of 
mathematical sciences v. III, Springer-Verlag, Berlin, 1988).
\bibitem{mel} V.K. Melnikov, Trans. Moscow Math. Soc. 12, 1 (1983)
\bibitem{hol} P. Holmes, Phys. Rep. 193, 137 (1990).
\bibitem{ivano} J. Koiller, J.R.T. de Mello Neto and I. Dami\~ao Soares, 
Phys. Lett. 110A, 260 (1985).
   \bibitem{letvie1}P.S. Letelier
and W. M. Vieira,  ``Chaos in periodically perturbed 
monopole+quadrupole potentials", Preprint UNICAMP-IMECC/1996.
\bibitem{xan}B.C. Xanthopoulos, Proc. Roy. Soc. A 378, 73 (1981).
\bibitem{chandra} S. Chandrasekhar, ``The mathematical theory of
 black holes", (Clarendon Press, Oxford, 1983).
\bibitem{frie} J.L. Friedman, Proc. Roy. Soc. A 335, 163 (1973).
\end{thebibliography}
\end{document}